\documentclass{article}
\usepackage{spconf,amsmath,graphicx}
\usepackage{stfloats}
\usepackage{url}
\usepackage[numbers,sort&compress]{natbib}
\title{LSSED: A Large-Scale Dataset and Benchmark for Speech Emotion Recognition}
\name{Weiquan Fan$^{1}$ 
    \qquad Xiangmin Xu$^{1}$ 
    \qquad Xiaofen Xing$^{1}$
    \qquad Weidong Chen$^{1}$ 
    \qquad Dongyan Huang$^{2}$
    \thanks{Xiaofen Xing is the corresponding author.}
    \thanks{Thanks to Datatang for support.}}
\address{$^{\star}$ School of Electronic and Information Engineering, South China University of Technology, China  \\ 
        $^{\dagger}$ UBTECH Robotics Corp, China}
\begin{document}
%
\maketitle
\begin{abstract}
Speech emotion recognition is  a vital  contributor to the next generation of human-computer interaction (HCI).
However, current existing small-scale databases have limited the development of related research.
In this paper, we present LSSED, a challenging large-scale english speech emotion dataset, which has data collected from 820 subjects to simulate real-world distribution.
In addition, we release some pre-trained models based on LSSED, which can not only promote the development of speech emotion recognition, but can also be transferred to related downstream tasks such as mental health analysis where data is extremely difficult to collect.
Finally, our experiments show the necessity of large-scale datasets and the effectiveness of pre-trained models.
The dateset will be released on \url{https://github.com/tobefans/LSSED}.

\end{abstract}
\section{Introduction}
\label{sec:intro}
Speech emotion recognition (SER) is a necessary part of the human-computer interaction system.
Although emotion itself is very abstract, it still has some obvious intonation characteristics.
Intuitively, sad voices are generally low-pitched and slow while happy voices are usually the opposite.
Up to now, many algorithms have emerged for existing dataset.

There are lots of researches carried out on SER.
In \cite{schuller2003hidden}, Schuller et al. applies continuous Hidden Markov Models (HMM) to introduce SER with a self-collected speech corpus.
Since 2004, some standardized speech emotion databases have been released. AIBO\cite{batliner2004you}, EMODB\cite{burkhardt2005database}, ENTERFACE\cite{martin2006enterface}, RML\cite{wang2008recognizing}, IEMOCAP\cite{busso2008iemocap}, AFEW\cite{dhall2012collecting}, and MELD\cite{poria2018meld}.
Among them, IEMOCAP\cite{busso2008iemocap} and MELD\cite{poria2018meld} are the databases with the most data.
IEMOCAP\cite{busso2008iemocap} collects 7,433 sentences (13 hours and 40 minutes in total) spoken by 10 people. MELD\cite{poria2018meld} contains 13,708 sentences (about 12 hours) from 407 people.
In \cite{lee2011emotion}, decision tree is utilized to mitigate error propagation on AIBO\cite{batliner2004you} and IEMOCAP\cite{busso2008iemocap}.
In \cite{stuhlsatz2011deep}, RBM is applied to learn discriminatory features on EMODB\cite{burkhardt2005database} and ENTERFACE\cite{martin2006enterface}.

With the rapid development of deep learning, Zhang et al. \cite{zhang2017speech} utilizes  DCNN to bridge the affective gap in speech signals on EMODB\cite{burkhardt2005database}, RML\cite{wang2008recognizing}, ENTERFACE\cite{martin2006enterface}.
At the same time, Satt et al. \cite{satt2017efficient} presents a system based on an end-to-end LSTM-CNN with raw spectrograms on IEMOCAP\cite{busso2008iemocap}.
Recently, Yeh et al.  \cite{yeh2020dialogical} proposes a dialogical emotion decoding algorithm to consecutively decode the emotion states of each utterance on IEMOCAP\cite{busso2008iemocap} and MELD\cite{poria2018meld}.

Although there have been certain level of  progression on SER, there is still a potentially serious overfitting problem, which may limit the development of SER.
As shown in \cite{zong2016cross, abdelwahab2018domain, liu2018unsupervised, luo2020nonnegative}, even if a high accuracy is achieved on a certain database, their performance may be poor when transferring to another database.
This is because the existing databases are generally small in scale, resulting in insufficient diversity, which is far from the real-world scenarios thus leading to the tendency of  model overfitting.
Therefore, a large-scale emotion dataset that can more comprehensively represent the real distribution is urgently needed to improve the generalization of existing algorithms.

Generally speaking, transfer learning can to a certain extent improve the performance of an algorithm.
Boigne et al. \cite{boigne2020recognizing} points out task-related transfer learning of recognizing emotions on small datasets.
For emotion recognition related task, a good pre-trained model is urgent since data collection is very difficult.
Taking the depression detection task as an example, there are only about a hundred subjects at most till date.
In our opinion, the pre-trained model from the SER task is more suitable for detecting depression, since it is more inclined to obtain acoustic features while the model from ASR task is prone to extract linguistic features.


In this paper, we present LSSED, a challenging large-scale english dataset for speech emotion recognition.
It contains 147,025 sentences (206 hours and 25 minutes in total) spoken by 820 people.
Based on our dataset, we can simulate a more comprehensive and rich data distribution of real-world scenarios so that deep neural networks can better model their distribution. 
Furthermore, since there is currently no non-semantic large-scale pre-training model, we release some pre-trained models with speech emotion recognition task.

\section{LSSED}
\label{sec:format}
In this section, we introduce our dataset, LSSED in details. 
LSSED collects a total of 147,025 utterances from 820 subjects, with an average duration of 5.05s.
As shown in Table \ref{tab:dataset}, the data volume of LSSED is very large, and its total duration (over 200 hours) can reach dozens of times than existing databases.

\begin{table*}[htbp]
  \caption{Comparison to existing public speech emotion datasets.}
  \label{tab:dataset}
  \resizebox{\linewidth}{!}{
  \centering
  \begin{tabular}{ lccccccc }
    \hline
    \textbf{Corpus} & \textbf{Published Time} & \textbf{Language} & \textbf{Speakers} & \textbf{Naturalness} & \textbf{h:mm} & \textbf{Number of sentences} & \textbf{Classes}\\
    \hline 
    AIBO\cite{batliner2004you}         & 2004 & Multiple   & 51 & Natural &   9:20    &   48401   &   11  \\
    EMODB\cite{burkhardt2005database}  & 2005 & German     & 10 & Acted   &   0:22    &   494     &   7   \\
    ENTERFACE\cite{martin2006enterface}& 2006 & English    & 43 & Acted   &   1:00    &   1170    &   6   \\
    RML\cite{wang2008recognizing}      & 2008 & Multiple   & 8  & Acted   &   0:42    &   500     &   6   \\
    IEMOCAP\cite{busso2008iemocap}     & 2008 & English    & 10 & Both    &   12:00   &   7433    &   10  \\
    AFEW\cite{dhall2012collecting}     & 2012 & Multiple   & 330 & Acted  &   unknown &   1426    &   7   \\
    MELD\cite{poria2018meld}           & 2018 & English    & 407 & Acted  &   13:40   &   13708   &   7   \\
    \hline
    LSSED                             & 2020 & English    &   820 &   Natural &   206:25  &   147025  &   11  \\
    \hline
    \end{tabular}}
\end{table*}

\subsection{Collection and Labeling}
The subjects that participate in the experiment are widely distributed with representations from both genders and variety of age groups.
Each subject would be recorded in one or several emotional videos sessions in an indoor lab environment with a camera pointing at him or her. 
In the video, the subject is induced by random questions as their utterances are associated with an emotional label.
The total length of a video is about 10-20 minutes. 

The utterances in each video dialogue are annotated by a professional annotation team.
Each utterance is annotated with the corresponding emotion label, including anger, happiness, sadness, disappointment, boredom, disgust, excitement, fear, surprise, normal, and other.
Note that some utterances in the video contain two or more emotions. 
In addition, each utterance is also annotated with auxiliary information, including the gender and age of the subject.

\subsection{Data Distribution}
As mentioned above, our database covers various groups of people.
Table \ref{tab:dis_gend_age} shows the conditional and joint distribution of the ages and genders.
In LSSED, the gender distribution is relatively balanced. The age distribution however has fewer elderly people.

\begin{table}[htbp]
  \caption{Data distribution for gender and age.}
  \label{tab:dis_gend_age}
  \centering
  \begin{tabular}{ lccc|c }
    \hline
    \textbf{} & \textbf{Young} & \textbf{Middle-aged} & \textbf{Old} & \textbf{Total}\\
    \hline
    \textbf{Female} &   253 &   167 &   65  &   485  \\
    \textbf{Male}   &   155 &   141 &   39  &   335  \\
    \hline
    \textbf{Total}  &   408 &   308 &   104 &   820  \\
    \hline
    \end{tabular}
\end{table}

\begin{figure}[t]
  \centering
  \includegraphics[width=\linewidth]{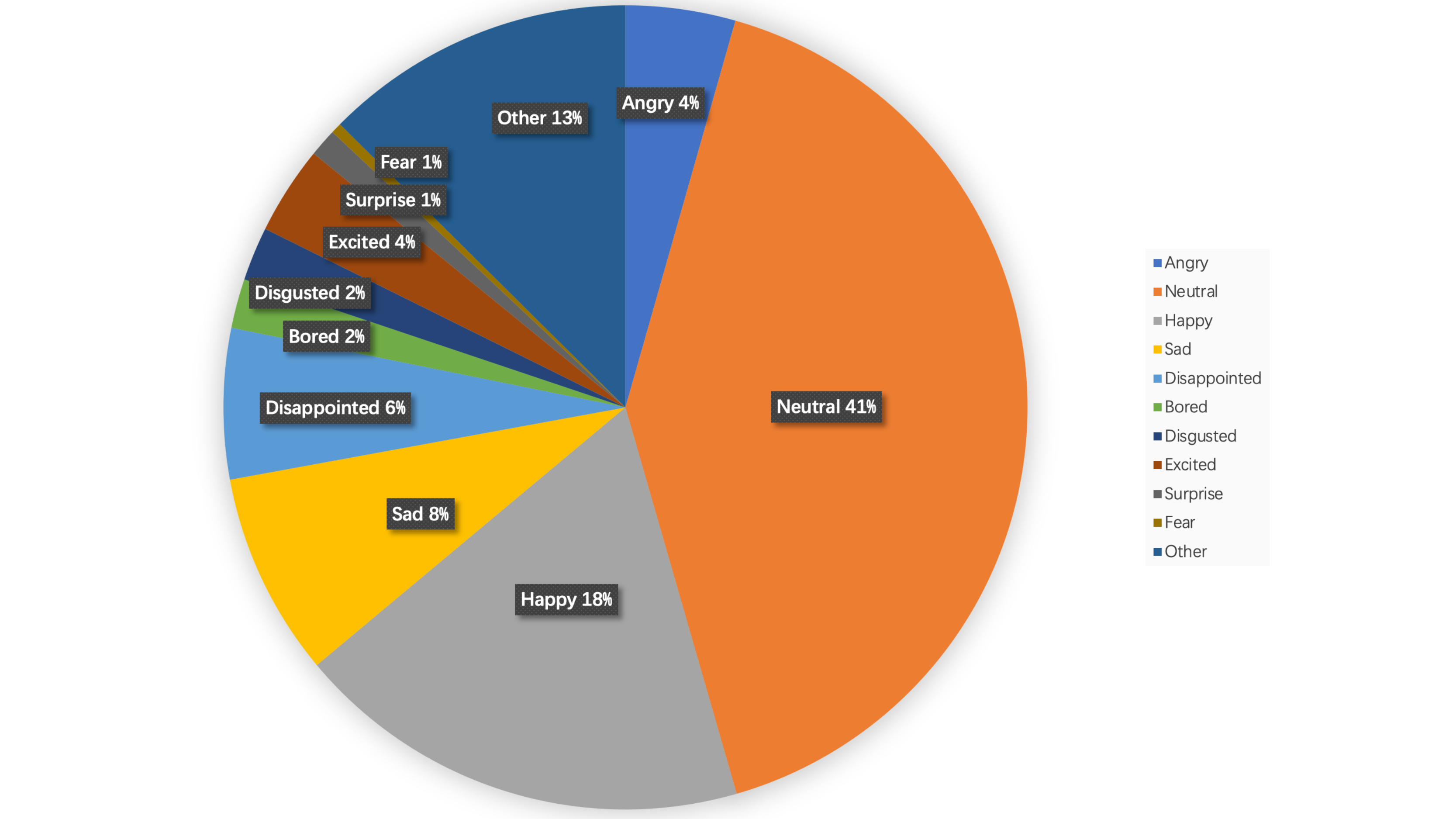}
  \caption{Distribution of data for each emotion labels}
  \label{fig:dis_class}
\end{figure}

A pie chart of the distribution of data for emotion labels is shown in Fig \ref{fig:dis_class}.
Since the subjects speak in a spontaneous environment, the more common neutral samples accounted for a larger proportion.
Next is happy, sad, disappointed, excited, and angry samples respectively.
The samples of these six categories account for 81\% of the total sample.
Then, the samples of boring, disgusting, fearful, and surprised are fewer, accounting for only 6\%.
In addition, 13\% of other uncommon samples can be used for tasks to distinguish whether they are common emotions.

In order to standardize future training benchmarks, we divided our LSSED dataset into training and test sets.
Specifically, we first shuffle the order of all samples, then set 20\% of the samples as the test set, and the rest as the training set.
It should be noted that we ensure the distribution of each emotion category in the training set and test set are the same or at least similar.
Table \ref{tab:dis_class} shows the specific distribution of data for emotion labels in the training set and test set respectively.

\begin{table}[htbp]
  \caption{Distribution of data for emotion labels in the subsets.}
  \label{tab:dis_class}
  \centering
  \begin{tabular}{ lcc|c }
    \hline
    \textbf{} & \textbf{Training} & \textbf{Test} & \textbf{Total}\\
    \hline
    \textbf{Angry}          &   1298    &   5192    & 6490 \\
    \textbf{Neutral}        &   12092   &   48369   & 60461 \\
    \textbf{Happy}          &  5406     &   21627   &  27033   \\
    \textbf{Sad}            &  2410     &   9641    &  12051   \\
    \textbf{Disappointed}   &  1781     &   7124    & 8905    \\
    \textbf{Bored}          & 583       &   2333    & 2916    \\
    \textbf{Disgusted}      &  636      &   2543    &  3179   \\
    \textbf{Excited}        &  1046     &   4182    &  5228   \\
    \textbf{Surprise}       &  331      &   1325    &  1656   \\
    \textbf{Fear}           &  126      &   502     & 628    \\
    \textbf{Other}          &  3696     &   14782   & 18478    \\
    \hline
    \textbf{Total}  &   \textbf{117620} &   \textbf{29405} &   \textbf{147025}  \\
    \hline
    \end{tabular}
\end{table}

\subsection{Preprocessing and feature extraction}
After obtaining the videos, we then convert them into audio signals at a sampling rate of 16kHz.
According to the start time and end time of each utterance, we cut out 147,025 audio utterances.
For each sentence, we use spectral subtraction algorithm \cite{upadhyay2015speech} to perform audio denoising. It subtracts noise on the short-time spectrum and then restores the audio.
Next, we increase the audio volume by a factor of 2 to make the sound louder.

After preprocessing, we perform STFT with Hann window length of 1024 points and the window shift of 512 points. A square operation follows to obtain the power spectrum.
The power spectrum is then passed through a triangular filter bank with 128 Mel-scales to simulate the human auditory perception system.

\subsection{Pre-trained Models}

We firstly select VGG \cite{simonyan2014very} and ResNet \cite{he2016deep} for pre-training, which are useful in many scenarios.
VGG builds a unified and simple structure to deepen the network, while ResNet proposes residual learning to ease the training procedure.

In order to better adapt to the specificity of speech, we propose PyResNet, an improved model of ResNet \cite{he2016deep}.
Due to the sufficient amount of data, PyResNet is based on ResNet50, ResNet101 or ResNet152.

Specifically, the second convolution layer in each layer of  ResNet is replaced with a pyramid convolution \cite{duta2020pyramidal} that can capture multi-scale information to solve the problem of uncertain time position of valid speech information.
In addition, we replaced the GAP layer with average pooling layer only in the time dimension to make the model insensitive to time and preserve the frequency information.




\section{Dataset Experiments}
\label{sec:experiments}


\subsection{Effectiveness of LSSED}
Although the current algorithms have achieved good results on many small-scale datasets, pre-trained models often cannot be well generalized to other datasets. This triggered our thinking about the scale of databases resulting in the collection and building of a large amount of database, that can be informative enough to train a model with good generalization.

In order to verify the effectiveness of different-scale datasets, we calculate the performance degradation based on ResNet152 as shown in Table \ref{tab:results_3}.
As indicated, the performance degradation is very large when the model trained from small-scale IEMOCAP\cite{busso2008iemocap} is tested on large-scale LSSED, while it is less when tested from large-scale to small-scale. 
This demonstrates the effectiveness of LSSED, since it simulates the real-world distribution.

\begin{table}[htbp]
  \caption{Performance degradation when testing in the target database compared with the source database (training in the source database).}
  \label{tab:results_3}
  \centering
  \begin{tabular}{ ll|cc }
    \hline
    \textbf{Source} & \textbf{Target} & \textbf{-WA} & \textbf{-UA}\\
    \hline
    IEMOCAP  &  LSSED      &   0.596  &   0.342  \\
    LSSED   &  IEMOCAP     &   \textbf{0.119}   &   \textbf{0.071}  \\
    \hline
    \end{tabular}
\end{table}

\subsection{Speech Emotion Recognition Benchmark}
We investigate some papers \cite{zhang2018attention, guizzo2020multi, ren2020generating} with open source code from recent SER papers.
Also, we carry out a series of contrast experiments based on commonly used backbone models, including VGG and ResNet.
In addition, we also test our PyResNet model mentioned in Section 2.4.

In the experiments, all algorithms use the training set and test set from LSSED.
The models of existing algorithms are based on the configuration in the original papers.
Our PyResNet and the backbone models are iterated for 60 epochs with batch size of 256 through the SGD optimizer with a weight decay of 0.001.
The learning rate (initialized to 0.01) drops to 10\% of the original every 20 epochs.
Consistent with the current mainstream SER experiments, we use four emotion categories, including angry, neutral, happy and sad.

\begin{table}[htbp]
  \caption{The performance of different methods on LSSED.}
  \label{tab:results_1}
  \centering
  \begin{tabular}{ lccc }
    \hline
    \textbf{Algorithm} & \textbf{Backbone} & \textbf{WA}  & \textbf{UA}\\ 
    \hline
    FCN-Attention\cite{zhang2018attention}  &    ALEXNet  &   0.570  &  0.250   \\
    MTS-3branches\cite{guizzo2020multi}  &    ALEXNet  &   0.570  &  0.250   \\
    MTS-5branches\cite{guizzo2020multi}  &    ALEXNet  &   0.570  &  0.250   \\
    MTS-3branches\cite{guizzo2020multi}  &    ResNet152  &   0.585  &  0.296   \\
    MTS-5branches\cite{guizzo2020multi}  &    ResNet152  &   0.582  &  0.311   \\
    ADV-Real\cite{ren2020generating}   &    VGG16  &   0.570  &  0.250   \\
    ADV-Fake\cite{ren2020generating}   &    VGG16  &   0.570  &  0.250   \\    
    ADV-Real\cite{ren2020generating}   &    ResNet152  &   0.548  &  0.381   \\
    ADV-Fake\cite{ren2020generating}   &    ResNet152  &   0.453  &  0.339   \\
    \hline
    VGG\cite{simonyan2014very}          &    VGG11              &   0.595  &  0.337    \\
    VGG\cite{simonyan2014very}          &    VGG13              &   0.604  &  0.393    \\
    VGG\cite{simonyan2014very}          &    VGG16              &   0.585  &  0.313    \\
    VGG\cite{simonyan2014very}          &    VGG19              &   0.585  &  0.370    \\

    ResNet\cite{he2016deep}             &    ResNet18           &   0.594  &  0.382  \\
    ResNet\cite{he2016deep}             &    ResNet34           &   0.598  &  0.355  \\
    ResNet\cite{he2016deep}             &    ResNet50           &   0.587  &  0.377  \\
    ResNet\cite{he2016deep}             &    ResNet101          &   0.592  &  0.332  \\
    ResNet\cite{he2016deep}             &    ResNet152          &   0.601  &  0.396  \\
    \hline
    PyResNet                            &    ResNet50           &   \textbf{0.615}  &  \textbf{0.420}  \\
    PyResNet                            &    ResNet101          &   \textbf{0.616}  &  \textbf{0.428}  \\
    PyResNet                            &    ResNet152          &   \textbf{0.624}  &  \textbf{0.429}  \\
    \hline
    \end{tabular}
\end{table}

The results are shown in table \ref{tab:results_1}.
This shows that the performance of existing algorithms on large-scale LSSED is not satisfactory.
More importantly, the accuracy (weighted and unweighted) of these algorithms is even lower than that of the basic VGG and ResNet models.

In addition, it is worth noting that our PyResNet achieves better results than the basic backbone models. This demonstrates that the improvement based on pyramid convolution is effective on large-scale database.
Since these algorithms are not excellent in overall performance on large-scale databases, it should be indicated that LSSED still has great challenges which means that speech emotion recognition is still a long way from being perfectly widely applicable.

\begin{figure}[htbp]
  \centering
  \includegraphics[width=\linewidth]{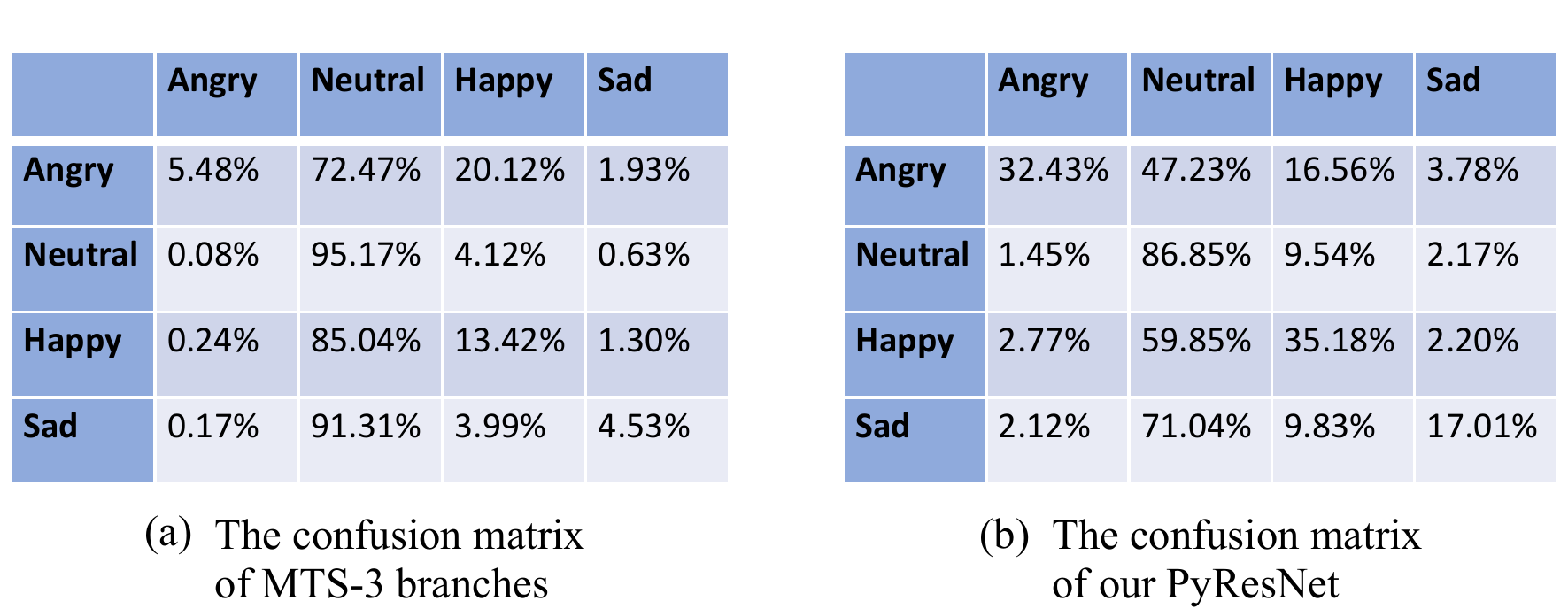}
  \caption{Confusion matrices of different algorithms. }
  \label{fig:confusion}
\end{figure}

Confusion matrices of both MTS-3 branches and PyResNet that use ResNet152 as the backbone is shown below.
Although they all use multi-scale convolution kernels, the former uses multi-scale kernels derived from one kernel, while the latter directly uses multiple different kernels with more powerful modeling capabilities.
As shown in Figure \ref{fig:confusion}, we can observe that neutral samples have a high probability of being correctly predicted, which is also the most common emotion.
But we should also note that both models have a prediction bias problem  for the neutral class.
We speculate that this is because each individual has different neutral standards.
In our future work, we will also take into account the resting (neutral) state of each individual.
In comparison, our PyResNet has a significant improvement in the angry, happy, and sad categories which are less predictable.

\subsection{Pre-trained Model in Downstream Task}
With the above pre-trained models, we want to further explore its applicability to downstream tasks.
We choose speech-based depression detection as our downstream task.
Due to the high professional requirements, it is very difficult to collect data on patients with depression. This leads to the current unsatisfactory effect of automatic depression detection.
It is therefore a natural idea to use a pre-trained model with sufficient prior knowledge to improve the detection accuracy.

These series of experiments are carried out on the DAIC-WOZ depression database, which is a subset of the Distress Analysis Interview Corpus (DAIC) \cite{gratch2014distress}.
There are 107 subjects in the training set, 35 in the development set, and 47 in the test set.
Each subject will be interviewed by an animated virtual interviewer and recorded with video and audio equipments.
The data will be annotated with the start time, end time and depression (or not) of each sentence.

We choose SER task and ASR task for transfer.
Firstly, we need to get the pre-trained models.
For SER, we use the pre-trained PyResNet with ResNet152 as a backbone.
For ASR, we use ESPNet \cite{watanabe2018espnet}, which is an end-to-end encoder-decoder structure network.

\begin{table}[htbp]
  \caption{The performance of different pre-trained models on DAIC-WOZ.}
  \label{tab:results_2}
  \centering
  \begin{tabular}{ lcc }
    \hline
    \textbf{Algorithm} & \textbf{WA} & \textbf{UA}\\
    \hline
    ESPNet (ASR)        &   0.657  &   0.500  \\
    PyResNet  (SER)        &   \textbf{0.714}   &   \textbf{0.583}  \\
    \hline
    \end{tabular}
\end{table}

The results of the experiment are shown in Table \ref{tab:results_2}.
The performance of transfer based on SER is better than that based on ASR.
This is because the features extracted by ASR are bias towards semantics while the features extracted by SER are bias towards acoustics.
Depression detection pays more attention to acoustic features which has larger gaps with ASR tasks.
Therefore, the pre-trained model on SER with a smaller gap has better performance.

Moreover, we also considered the differences in bandwidth between SER and ASR when framing.
ASR generally uses a narrow window length of about 25ms. This means that it pays more attention to changes in time and has a higher time resolution.
For SER, we use a wide window length of about 65ms, which means that the frequency information in each frame is richer and the frequency resolution is higher.
In general, a high time resolution is conducive to extracting semantic features from frame by frame and a high frequency resolution is conducive to extracting acoustic features.
Therefore, for downstream tasks such as depression detection, the SER pre-trained model with high frequency resolution and smaller gap may be a better choice.

\section{Conclusion}
\label{sec:conclusion}
In this work, we present LSSED, a challenging large-scale english database for speech emotion recognition that can simulate real distribution. 
We point out that existing algorithms tend to overfit small-scale databases and thus cannot be well generalized to real scenes. 
Furthermore, we release some pre-trained models based on LSSED. 
These models can not only promote the development of SER, but can also be transferred to similar downstream tasks like mental health analysis where data is extremely difficult to collect.

\small
\bibliographystyle{IEEEbib}
\bibliography{refs}

\end{document}